\documentstyle[preprint,floats,tighten,prl,aps]{revtex}

\def\lesssim{\ \hbox{\raise 2pt \hbox{$<$} \kern -13pt
                     \lower 3pt \hbox{$\sim$}}\ }
\def\greatersim{\ \hbox{\raise 2pt \hbox{$>$} \kern -13pt
                     \lower 3pt \hbox{$\sim$}}\ }
\input epsf.tex
\def\desepsf(#1 width #2){\epsfxsize=#2 \epsfbox{#1}}

\begin{document}

\draft

\title{Power-like corrections and the determination of the\\ 
 gluon distribution}
\author{F.\ Hautmann}
\address{Institut f{\" u}r Theoretische Physik, 
Universit{\" a}t Regensburg, 93040  Germany}

\maketitle

\begin{abstract}
Power-suppressed corrections  
to parton evolution may affect the theoretical 
accuracy of current  determinations of 
parton distributions. 
We study the role of   multigluon-exchange terms 
in the extraction of the gluon distribution for the 
Large Hadron Collider (LHC).   
Working in the high-energy approximation, 
we analyze   multi-gluon  contributions  in powers of 
$1/Q^2$. 
We find a  moderate, negative      correction 
to  the structure function's derivative $d F_2 / d \ln Q^2$, 
characterized by a slow fall-off 
in the region of   low to medium $Q^2$    
relevant for determinations of  the gluon  at small 
momentum fractions. 
\end{abstract} 


The estimate of the theoretical accuracy  on 
 the determination of parton distributions  
is relevant for     phenomenology  at the 
 Large Hadron Collider (LHC)~\cite{thuncert,mrst04,cteq02}. 
A potential source of theory uncertainty 
 is given by  
power-suppressed contributions to parton evolution. 
 In particular,    the extraction of the gluon 
distribution  for small  momentum fractions 
depends on data in the region of high energies and 
 moderate to low $Q^2$,  in which power-like  
 corrections from 
  multiple parton scatterings are potentially 
   significant.

The purpose of this note is 
to derive an estimate  
of these corrections, based on 
  high-energy  amplitudes  for 
  multigluon exchange~\cite{muelect}. 
We concentrate on  the power  
correction $\Delta$  to the $Q^2$ derivative of the 
structure function $F_2$
\begin{equation} 
\label{df2} 
{{d F_2} \over { d \ln Q^2 }} = K 
\otimes G \ \left[ 1 + \Delta 
\right] + \ {\rm{quark}} \; \ {\rm{term}} \;  
\hspace*{0.2 cm}  , 
\end{equation} 
where $K$ is the perturbative kernel,  
 $G$ is the gluon distribution, and $\Delta$ 
 is ${\cal O} (1 /Q^2)$. 
 This correction contributes  to the theory 
uncertainty on  $G$, since 
 below $x \lesssim 10^{-2}$ this   
 is mostly determined 
from data for $d F_2 / d \ln Q^2$.

The theoretical framework  to treat  multiple 
scatterings is based on the 
$s$-channel picture of DIS~\cite{muelect}, and 
its basic 
 degrees of freedom  
 are 
described  
by matrix elements of  eikonal lines. 
The corresponding 
predictions, incorporating 
 nonperturbative dynamics,  are valid down to  small $Q^2$. 
To enforce  consistency  with the  framework of standard 
parton analyses~\cite{mrst04,cteq02,thorne05} 
we will expand the answer  in powers of $1 /Q^2$.  
We study the behavior of the 
power expansion at low $Q^2$ and $x$, and 
 identify the power corrections by 
subtraction of  the leading-power contribution. 

We find moderate, but non-negligible   negative 
 corrections,  
increasing in size as $x$ decreases. 
We find that 
with a physically natural choice of 
 parameters in the eikonal matrix elements 
we can achieve a sensible description of data 
 and still have  power corrections to 
$d F_2 / d \ln Q^2$ 
that do not exceed
 the leading power already at  
$Q^2 \greatersim 0.5 $ GeV$^2$. 
However we find that for  small $x$  
   the corrections 
have a slow fall-off  with $Q^2$ 
in the region of intermediate  $Q^2$,   
 $Q^2 \simeq 1 \div 10 $ GeV$^2$, 
 behaving effectively like  $1/Q^\lambda$ 
in this region, 
with  
$   \lambda $ close to 1 for $x \lesssim 10^{-3}$.  
This behavior results from summing
the power expansion. 
As a consequence   
the power corrections  stay larger than 
$ 10 \%$ up  to $Q^2$ of a few GeV$^2$ for $x$ 
below $10^{-3}$.

The contents of the paper is as follows.
We first give the $s$-channel 
results  that  provide  
the   basic elements 
 for the evaluation of the power 
expansion. 
Then we focus  on the $Q^2$ derivative of the 
structure function.  We examine  the 
next-to-leading-power contribution and the sum of the
 power series,   and present numerical results.

We begin by recasting the result~\cite{muelect} 
for multi-gluon contributions to DIS structure functions 
in the   Mellin-transform  representation. This is 
convenient to analyze  the   $1/Q^2$ expansion. 
  In the high-energy approximation 
 we describe gluon exchange   
in terms of  
eikonal-line operators 
\begin{equation}
\label{eikF}
V({\bf z}) = {\cal P}\exp\left\{
-ig_s \int_{-\infty}^{+\infty}dz^- 
{ A}^+_a(0,z^-,{\bf z})t_a
\right\} \hspace*{0.4 cm}  , 
\end{equation}  
where $A$ is the color potential and 
${\cal P}$ is the path-ordering. 
Following~\cite{hs00} we notate the 
  matrix elements of eikonal operators 
at transverse positions ${\bf b}$ 
 and ${\bf b} + {\bf z}$ as 
\vspace{1cm}
\begin{equation}
\label{Xidefin}
\Xi( {\bf z}, {\bf b}) = 
\int [ d P^\prime ] \ 
\langle P'|\frac{1}{N_c} \ {\rm Tr}\{1-
V^{\dagger}( {\bf b} + {\bf z})\,
V({\bf b})
\} |P \rangle \hspace*{0.4 cm} .  
\end{equation}  
Here  $ [ d P^\prime ] = dP^{\prime +} d^2 {\bf P}^\prime/ 2 
 P^{\prime +} (2\pi)^{3}$, 
$ {\bf z}$ is 
the  transverse separation between 
the eikonal lines, and $ {\bf b}$ is the impact parameter. 
 The result~\cite{muelect} for the transverse structure 
function $F_T$ can be written  
  as 
\begin{equation} 
\label{urepr}
x F_T (x , Q^2) =  
 \int_{c - i \infty}^{c + i \infty}  
\frac{du}{2 \pi i}   \int d^2 {\bf b} 
\ {\widetilde \Xi} ( u, {\bf b} ) 
\ \Phi (u) \;\; , 
\end{equation}
where $0 < c < 1$, ${\widetilde \Xi}$ is the Mellin 
transform of $\Xi$ 
\begin{equation}
\label{Xitilde}
{\widetilde \Xi} ( u, {\bf b} ) = 
\int\! \frac{d^2{\bf z}}{ \pi {\bf z}^2} \
( {\bf z}^2  )^{u - 1} \ \Xi ( {\bf z}, {\bf b}) \;\; ,
\end{equation}
and $\Phi$ is a calculable coefficient. To lowest order  
\begin{equation} 
\label{Phiu}
\Phi (u) = 
\sum_a e_a^2 \frac{N_c }{ 16 \pi^{4} } 
 {1 \over u}  
( Q^2  )^{u}  \frac{\pi^2}{ 4^u} 
\frac{ \Gamma(3-u) \Gamma(2-u) 
\Gamma(1-u) \Gamma(2+u) }{\Gamma(5/2-u) \Gamma(3/2+u)} 
\;\;\;\;  ,  
\end{equation}
where $N_c = 3$,  and $e_a$ is the electric charge 
of  quark of type $a$. 

From this approach the parton-model framework 
is recovered 
through an expansion 
in powers of $g A$, valid for small ${\bf z}$. 
In particular, the coefficient of the quadratic term 
in the expansion of $\Xi$  can be  related 
to the  gluon distribution $ G$~\cite{muelect},    
\begin{equation}
\label{fxirel}
\int d^2 {\bf b} \ {\Xi} 
 \simeq  {{ \pi g_s^2 T_R } \over {4 N_c} } \  
  (x G)  \  {\bf z}^2 \ \left( 1 + 
  {\cal O} (|{\bf z}|) \right) 
\hspace*{0.4 cm}  ,  
\end{equation} 
where $T_R = 1/2$.

 The quark distribution can be dealt with 
 by the same method. 
 The main difference is that while 
 in the structure 
 function case the ultraviolet region is 
 naturally regulated by the physical 
  scale $Q^2$, in the case of the 
  quark distribution we need 
   to treat  the ultraviolet 
  divergences.   
  The result in dimensional regularization 
 is~\cite{hsprep}   
\begin{equation}
\label{fqresult}
x q (x,\mu) = 
(\mu^2)^{- 2 \epsilon} \ \int d^{2 - 2 \epsilon} 
{\bf z} \ d^{2 - 2 \epsilon}{\bf b}
\ w({\bf z}) \
\Xi({\bf z}, {\bf b}) - {\rm{UV}} \hspace*{0.4 cm} , 
\end{equation}
where  
\begin{equation}
\label{wresult}
w({\bf z}) =
{N_c \over 3 \pi^4 }\, \frac{1}{{\bf z}^4} \
\left(\frac{\mu^2 {\bf z}^2}{4\pi}\right)^{2\epsilon}
\frac{\Gamma(2-\epsilon)^2}{1-2\epsilon/3} 
\hspace*{0.4 cm}   , 
\end{equation}
and  we have indicated by UV  the 
ultraviolet subtraction. This is required since  
in Eq.~(\ref{Xidefin}) 
  $ (1 -  V^{\dagger} V) \to 0$   for
${\bf z} \to 0$  
and $ \Xi \propto {\bf z}^2$.  
Using Eqs.~(\ref{fxirel}),(\ref{wresult}) 
    the 
$x \ll 1$ form of the evolution 
equation for the quark  is reobtained from Eq.~(\ref{fqresult}). 
 Power-suppressed contributions   arise 
 from the difference between the  terms   left over
 from the ultraviolet regularization 
  in the quark-distribution and 
 structure-function case~\cite{hsprep}. 
 In general, these depend on the  scheme used for the 
ultraviolet subtraction. 
This dependence does not enter  in the 
 correction to the structure function's derivative   
which we consider next.

The matrix element $\Xi$  is 
nonperturbative and is to be determined from 
experiment. 
For the numerical calculations that follow we 
model its functional form  according to the 
 model~\cite{mue99,golecwue} 
\begin{equation}
\label{melgau}
{\widetilde \Xi} ( u , {\bf b} )= 
{{ \Gamma(u) } \over {1 -u} } 
\ \left( { {\mu_s^2 ( {\bf b} ) } 
\over 4} \right)^{1 - u} 
\;\;\;\;  , 
\end{equation}
where $\mu_s$ is the saturation scale, with  
 ${\bf b}$ dependence as in~\cite{mue99}. 
The  operator  relation  (\ref{fxirel}) 
implies 
for model  (\ref{melgau}) that 
\begin{equation}
\label{qsgen}
 \int d {\bf b} \ \mu_{s}^2  = 
   { 4 \pi^2 T_R \over N_c}  \ \alpha_s (\mu_r) 
   \  x G  (x , \mu_f)  \hspace*{0.2 cm}    .     
\end{equation}    
In Eq.~(\ref{qsgen}) we have indicated 
explicitly the dependence of the running coupling 
and gluon distribution on 
the renormalization/factorization 
scales $\mu_r$, $\mu_f$. In the present context 
the choice of these scales amounts to 
specifying  the model for $\Xi$.

\begin{figure}[htb]
\vspace{135mm}
\includegraphics{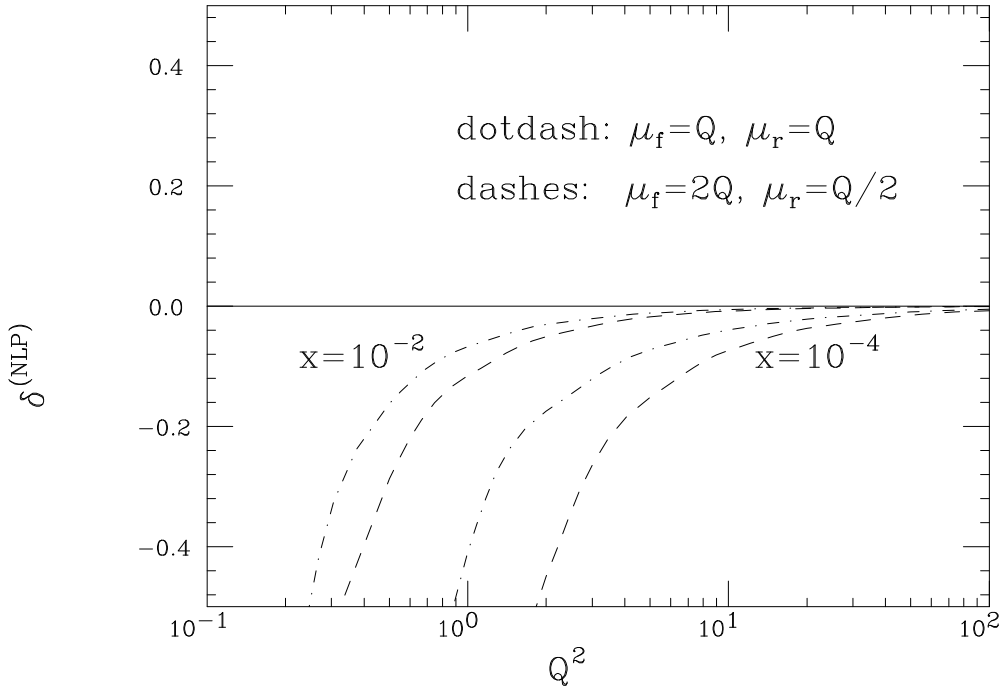}
\caption{Relative size of the next-to-leading- and 
 leading-pole contributions to  
the $Q^2$ derivative of the transverse structure 
function.}
\label{figdnlp}
\end{figure}

Consider now
the  derivative of 
$F_T$ with respect to $\ln Q^2$, 
$F^\prime_T \equiv d F_T / d \ln Q^2$. 
Taking the derivative   
 cancels the factor $1/u$ in Eq.~(\ref{Phiu}).  
 We determine the expansion of $F^\prime_T$ 
in powers of $1/Q^2$ 
by closing the integration contour 
in  the complex $u$-plane to the left and 
evaluating the  residues at the poles 
of the integrand.  
First we  verify that the result from 
the leading pole (LP) $u =0$  
coincides with that  from 
Eq.~(\ref{fqresult}) for the quark 
distribution.  We get  
\begin{equation} 
\label{leadpow}
x F^\prime_{T ,{\rm{LP}} }
=  \sum_a e_a^2 \frac{N_c }{ 3 \pi } \ 
\ \int d^2 {\bf b} \ 
{\rm{Res}}_{u = 0} \ {\widetilde \Xi} 
   \hspace*{0.2 cm}    .  
\end{equation}
By inserting  
Eqs.~(\ref{melgau}),(\ref{qsgen}), 
Eq.~(\ref{leadpow}) yields 
the perturbative leading-power coefficient.  
We identify the power-suppressed correction 
by subtracting off this contribution. 
Next we consider the contribution from the 
 next-to-leading pole (NLP) $u=-1$, 
$F^\prime_{T , {\rm{NLP}} }$. This is  proportional 
to the $u=-1$ residue of ${\widetilde \Xi}$.  
In Fig.~\ref{figdnlp} 
 we compute the ratio
\begin{equation} 
\label{deltanlp}
\delta^{({\rm{NLP}})}
\equiv   F^\prime_{T , {\rm{NLP}} } / 
( F^\prime_{T ,{\rm{LP}} } + F^\prime_{T , {\rm{NLP}}} ) 
   \hspace*{0.2 cm}     
\end{equation}
versus $Q^2$ at  
$x = 10^{-2}$ and $x = 10^{-4}$ 
for two different choices of 
 $\mu_r , \mu_f $. 
The natural scale for $\mu_r$ and $\mu_f$  should be  
 set by  the 
 inverse of the mean transverse distance ${\bf z}$. 
 For the  illustration 
 in Fig.~\ref{figdnlp} we 
 take this scale to be on the order of $Q$, and 
 plot results for 
$\mu_f = Q$, $\mu_r = Q$  
and 
$\mu_f = 2 Q$, $\mu_r = Q / 2$. 
We use the CTEQ   parton distributions~\cite{cteq02}.

 The ratio $\delta^{({\rm{NLP}})}$  goes like $1/Q^2$. 
We see from Fig.~\ref{figdnlp} that 
the size of the NLP contribution 
is rather sensitive 
to the scales $\mu_f$ and $\mu_r$. This is not 
surprising. The definition of the model for 
$\Xi$, embodied in the choice of  $\mu_f$, $\mu_r$, 
corresponds to defining the (otherwise arbitrary) 
separation of perturbative and nonperturbative 
effects. In particular, varying these scales 
amounts to effectively simulating contributions of 
higher perturbative order. The variation of the 
nonperturbative power correction in 
Fig.~\ref{figdnlp} says that this  is unambiguously 
defined only once we specify what we include in the 
perturbative part of the calculation. 
 See~\cite{gardiwei} for an extensive study of 
 the issue of  scale-setting and running coupling 
effects in high-energy evolution. In what follows 
we simply set  the scales from comparison with 
 experimental data.

Beyond the next-to-leading power, the poles in 
the $u$-plane have multiplicity 
higher than 1, leading to $\ln Q^2$ enhancements of 
the power corrections.  The 
 order-$n$ term is of the form 
\begin{equation}
\label{strucxi}
C(n, \ln Q^2) {{\xi_n } \over {(Q^2 )^n }}  \;\; , \;\; 
\end{equation}
where 
 $\xi_n$ give the dimensionful nonperturbative scale  
 in terms of the ${\bf b}$-integral  
of  the moments  (\ref{Xitilde}) of $\Xi$, 
and $C$ are  coefficients 
determined from Eq.~(\ref{Phiu}). 
 Through the moments of $\Xi$ 
 the correction (\ref{strucxi}) receives 
contribution from the  exchange of 
any number of gluons via eikonal operators.

\begin{figure}[htb]
\vspace{125mm}
\includegraphics{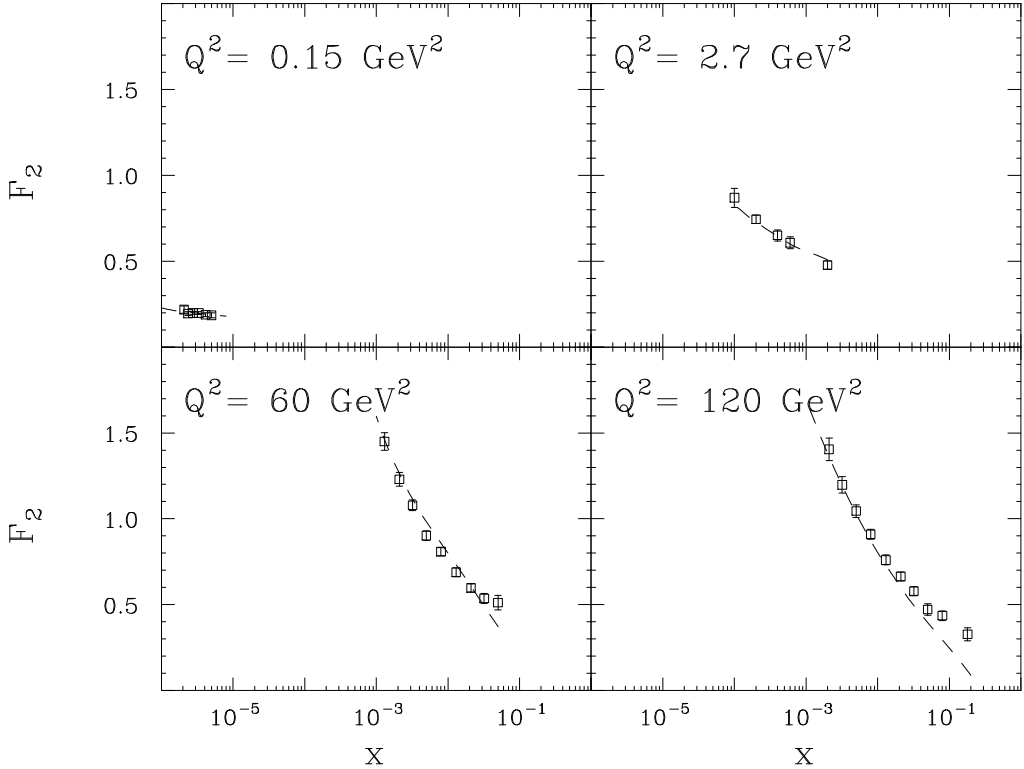}
\caption{Predictions for the structure 
function $F_2$ compared with  the data~[11].}
\label{figf2data}
\end{figure}

We now proceed to evaluate numerically the 
contribution of all powers in $1/Q^2$. 
We go  to ${\bf z}$-space via the inverse 
 Mellin transform  and 
let   the 
 scales $\mu_f$ and $\mu_r$ 
 vary with the distance ${\bf z}$. 
 Details for the numerical integration 
 in coordinate
 space are given 
 in \cite{hsprep}. Adding to Eq.~(\ref{urepr}) 
the contribution of  longitudinal $F_L$,  
 we compute the structure function $F_2$ and 
 tune 
 the factorization/renormalization scales, 
 $\mu_f= c_1 / |{\bf z}|$  and 
$\mu_r= c_2 / |{\bf z}|$,  by comparing the answer 
with the 
experimental data~\cite{zdata}.  
The result of doing this 
 is reported in Fig.~\ref{figf2data}, where 
 $c_1 = 4$, $c_2 = 0.32$.

\begin{figure}[htb]
\vspace{115mm}
\includegraphics{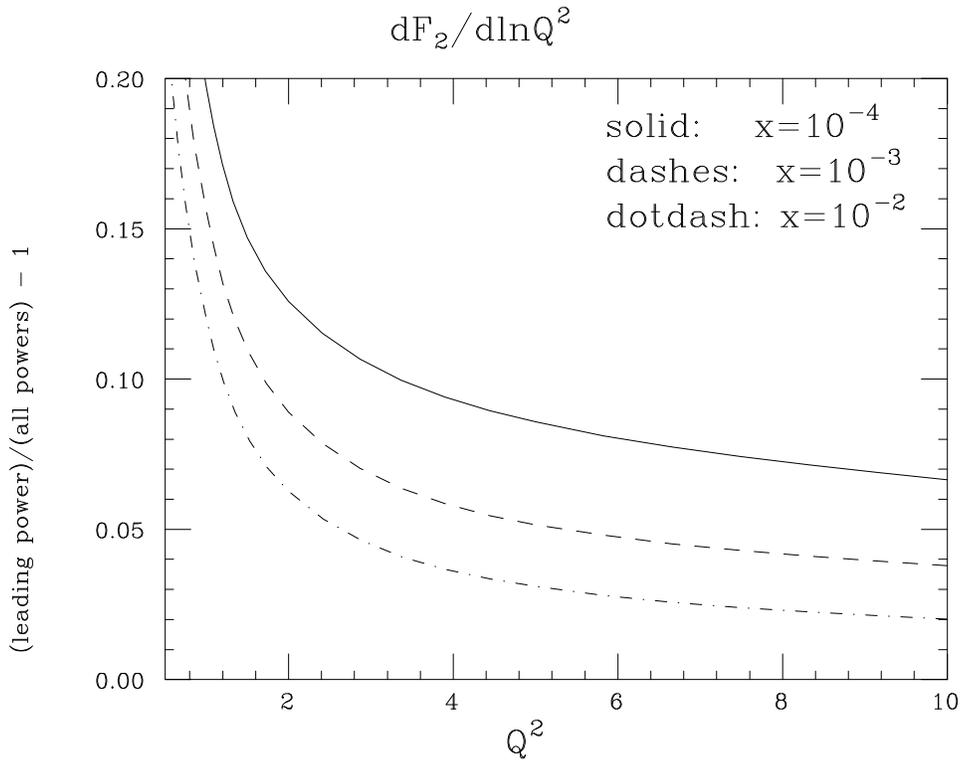}
\caption{Power corrections to 
 $d F_2/d \ln Q^2$ versus $Q^2$ at different
 values of $x$.}
\label{figallpwr}
\end{figure}

Using these values for the model parameters, 
we turn to the derivative  $d F_2 / d \ln Q^2$. 
We calculate the power correction by subtraction 
of the leading power as described around  
Eq.~(\ref{leadpow}).  
In  Fig.~\ref{figallpwr} we 
 plot  the 
result for the  power correction  
normalized to the  full answer and multiplied 
by $(-1)$. 
We see from  Fig.~\ref{figallpwr}  that above 
$Q^2 = 1$ GeV$^2$  the corrections are below 
 $ 20 \%$  for $x \greatersim 10^{-4}$. We  take 
this as an indication  that  
the power expansion is not breaking down,  
at least to such values of $x$. 
In fact, we find that 
corrections do not exceed the leading power already at 
$Q^2 \simeq 0.5$ GeV$^2$ for $x$ above $10^{-5}$.

Fig.~\ref{figallpwr} also shows, however,  that 
as $x$ decreases 
the power corrections remain non-negligible up to 
higher and higher $Q^2$. For $x \lesssim 10^{-3}$ 
it takes $Q^2$ of a few GeV$^2$ before the correction 
 is less than  $ 10 \%$. 
This  is associated 
with  the curves  
having a  rather  slow decrease   in the range of 
medium $Q^2$  in the figure, much slower than the 
asymptotic $1 / Q^2$.  For instance, 
for $x \simeq 10^{ - 3}$ 
the behavior in the 
region $Q^2 \simeq 1 \div 10 $ GeV$^2$ is 
closer to an effective power 
$1 / Q^\lambda$ with $\lambda \simeq 1.2$. 
This slow  fall-off   results from 
summing the  terms (\ref{strucxi}). 
We leave to a future study the question of 
whether  
this behavior can be interpreted in   
 terms  of   
  an effective, $x$-dependent semihard 
scale 
on the order of the   GeV.

In summary,  
the results above indicate  that 
the power expansion should still work at 
   $x$ and $Q^2$ as low as in the region  
of the data presently used 
for determinations of the gluon distribution $G$ 
for the LHC, 
 but subleading  corrections to 
$d F_2 / d \ln Q^2$ from multigluon exchange 
are non-negligible in this region and 
  contribute  to the 
theory uncertainty on  $G$. 
We have also computed the analogous corrections 
for the 
  derivative of the 
transverse structure function~\cite{hsprep}, and 
we find that 
these are 
generally smaller than in the case of $F_2$. 
 This   may be 
regarded 
as    yet another motivation 
for the importance of a separate measurement of the 
longitudinal component $F_L$~\cite{durham06}.  

A word of caution is needed  in interpreting these 
 results. 
Multigluon amplitudes are treated in the 
 high-energy approximation. 
Also, the  modeling of the nonperturbative 
 matrix elements  
and the summation of  
 the power series expansion  
 call for  a firmer  understanding. 
Besides, 
power corrections  from sources 
other than  
that considered here may  be relevant as well. 
In particular,   corrections  
 from self-energy graph insertions   
are still largely unexplored 
for  flavor-singlet observables~\cite{bubblex}.

Nevertheless, the method presented above 
allows one to obtain 
an estimate of multi-gluon corrections which 
can be made 
consistently with perturbative evolution  
order by order.  
It is based on subtraction of the leading 
 pole in Mellin space. This serves to specify 
the definition of the power correction. 
In this work we have been concerned  with 
 the 
contribution to  the $Q^2$ 
derivative of the structure function, 
relevant for the extraction of $G$. But 
the approach can in principle  be extended 
to evaluate   corrections to $F_2$ itself, and 
to processes  directly  coupled to gluons.  The latter  
will be especially interesting  for studying 
multiple-scattering effects  in  the production 
 of jet final states. 

\vskip 1 cm

\noindent 
{\bf Acknowledgments}. I thank D.~Soper for 
collaboration on topics related to this work.

\end{document}